\documentclass[a4paper,11pt]{article}
\pdfoutput=1 

\usepackage{jinstpub} 

\usepackage{cleveref}
\usepackage{placeins}
\usepackage{caption}
\usepackage{subcaption}

\newcommand{\nueb}{\ensuremath{\bar{\nu}_e}}

\title{\boldmath The SoLid anti-neutrino detector's readout system}

\author[a]{L. Arnold}
\author[b]{W. Beaumont}
\author[a,1]{D. Cussans\note{Corresponding author.}}
\author[a]{D. Newbold}
\author[c]{N. Ryder}
\author[c]{and A. Weber}

\affiliation[a]{University of Bristol}
\affiliation[b]{University of Antwerp}
\affiliation[c]{University of Oxford}

\emailAdd{David.Cussans@bristol.ac.uk}

\abstract{
    The SoLid collaboration have developed an intelligent readout system to reduce their 3200 silicon photomultiplier detector's data rate by a factor of 10000 whilst maintaining high efficiency for storing data from anti-neutrino interactions.
    The system employs an FPGA-level waveform characterisation to trigger on neutron signals.
    Following a trigger, data from a space-time region of interest around the neutron will be read out using the IPbus protocol.
    In these proceedings the design of the readout system is explained and results showing the performance of a prototype version of the system are presented.
}


\collaboration[c]{on behalf of the SoLid collaboration}

\proceeding{
    Topical Workshop on Electronics for Particle Physics (TWEPP2016)\\
	Karlsruhe, Germany\\
	September 26th-28th, 2016
}

\usepackage{lineno}
	
\begin{document}


\maketitle
\flushbottom

\section{Introduction}
\label{sec:intro}


The reactor anti-neutrino anomaly \cite{ref:reactor_anomaly} and the gallium anomaly \cite{ref:gallium_anomaly} both occur due to a measured deficit of electron anti-neutrinos detected within 100~m of nuclear reactors or intense radioactive sources.
The SoLid collaboration aim to determine whether these anomalies may be due to a very short baseline oscillation to a new, sterile neutrino flavour.
The SoLid experiment will search for an oscillation by measuring the anti-neutrino energy spectrum at a range of distances between 5 and 10~m from the highly enriched uranium BR2 reactor core at SCK$\bullet$CEN.
Deploying a tonne scale detector at ground level, next to a nuclear reactor is particularly challenging due to the high rate of cosmic ray and reactor related background events.

The SoLid collaboration has developed a novel anti-neutrino detector to efficiently identify anti-neutrino events despite the high background rate, requiring an FPGA level trigger scheme designed specifically for collecting data from inverse beta decay events.
This paper describes the detector and the design of the readout hardware in sections \ref{sec:detector} and \ref{sec:hardware}.
A prototype version of the electronics, supporting only eight channels, was produced and tested.
Results demonstrating the performance of the prototype system are presented in section \ref{sec:8chan}.
The design of the firmware and neutron based trigger scheme are explained in section \ref{sec:firmware}.

\section{The SoLid Detector}
\label{sec:detector}


The SoLid experiment detects electron anti-neutrinos via the inverse beta decay (IBD) interaction, $\bar{\nu}_e + p \rightarrow e^+ + n$.
The detection signature is a signal from the $e^+$ followed by a signal from the capture of the thermalised neutron.
The kinetic energy of the positron produced by IBD is very strongly correlated with the energy of the parent \nueb{}~.
Hence measuring the positron energy spectrum measures the \nueb{}~spectrum.

\begin{figure}[htbp]
    \centering
    \begin{subfigure}{0.9\textwidth}
        \includegraphics[width=0.5\textwidth]{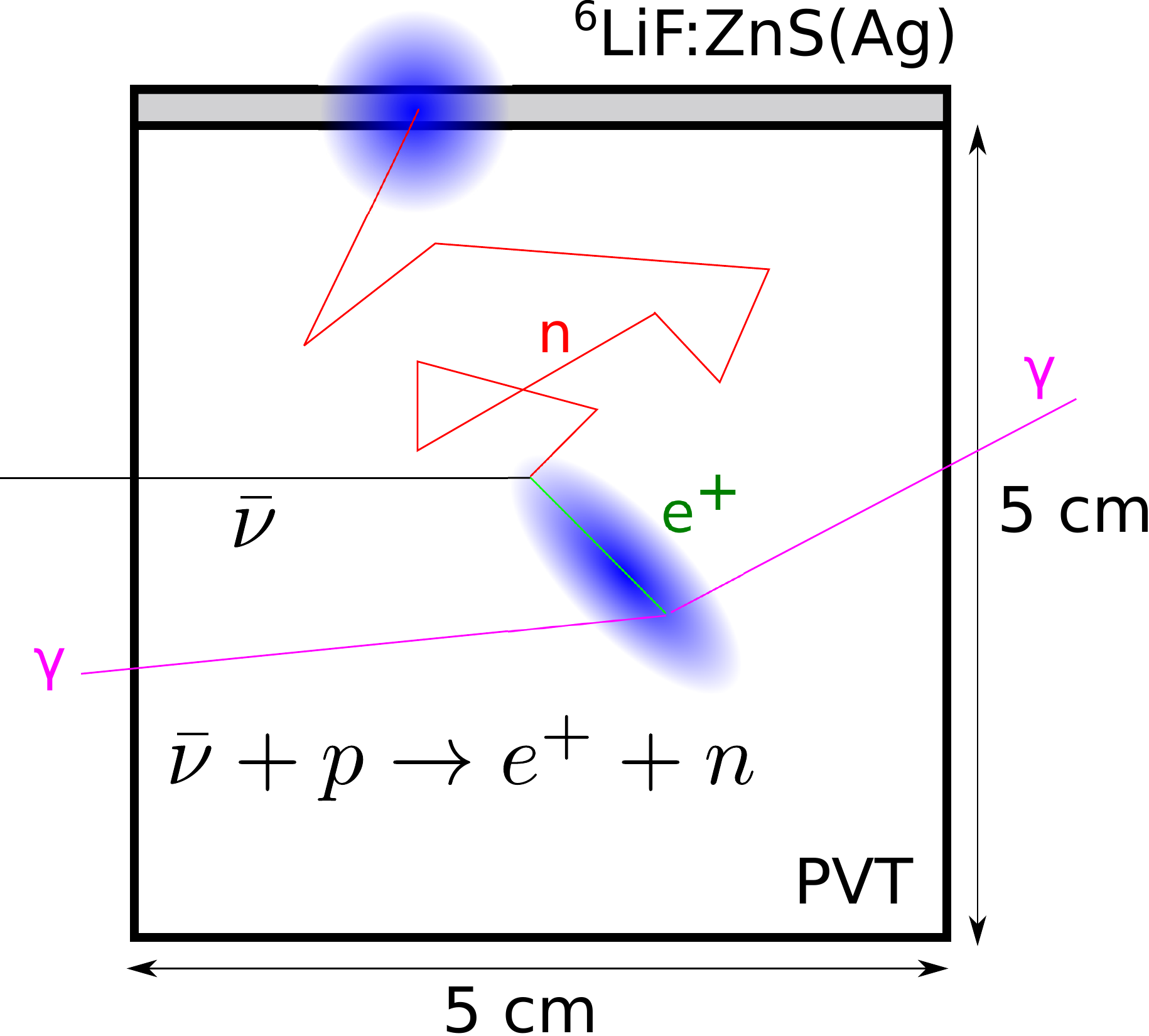}
        \caption{An electron anti-neutrino undergoes an inverse beta decay when interacting with a free proton in the PVT. A positron and neutron are emitted. The positron gives a prompt scintillation signal in the PVT. The neutron thermalises and is captured on $^6$Li resulting in a delayed scintillation signal in the $^6$LiF:ZnS(Ag) screen.}
        \label{fig:compositescintillator}
    \end{subfigure}
    
    \begin{subfigure}{0.9\textwidth}
        \includegraphics[width=0.9\textwidth]{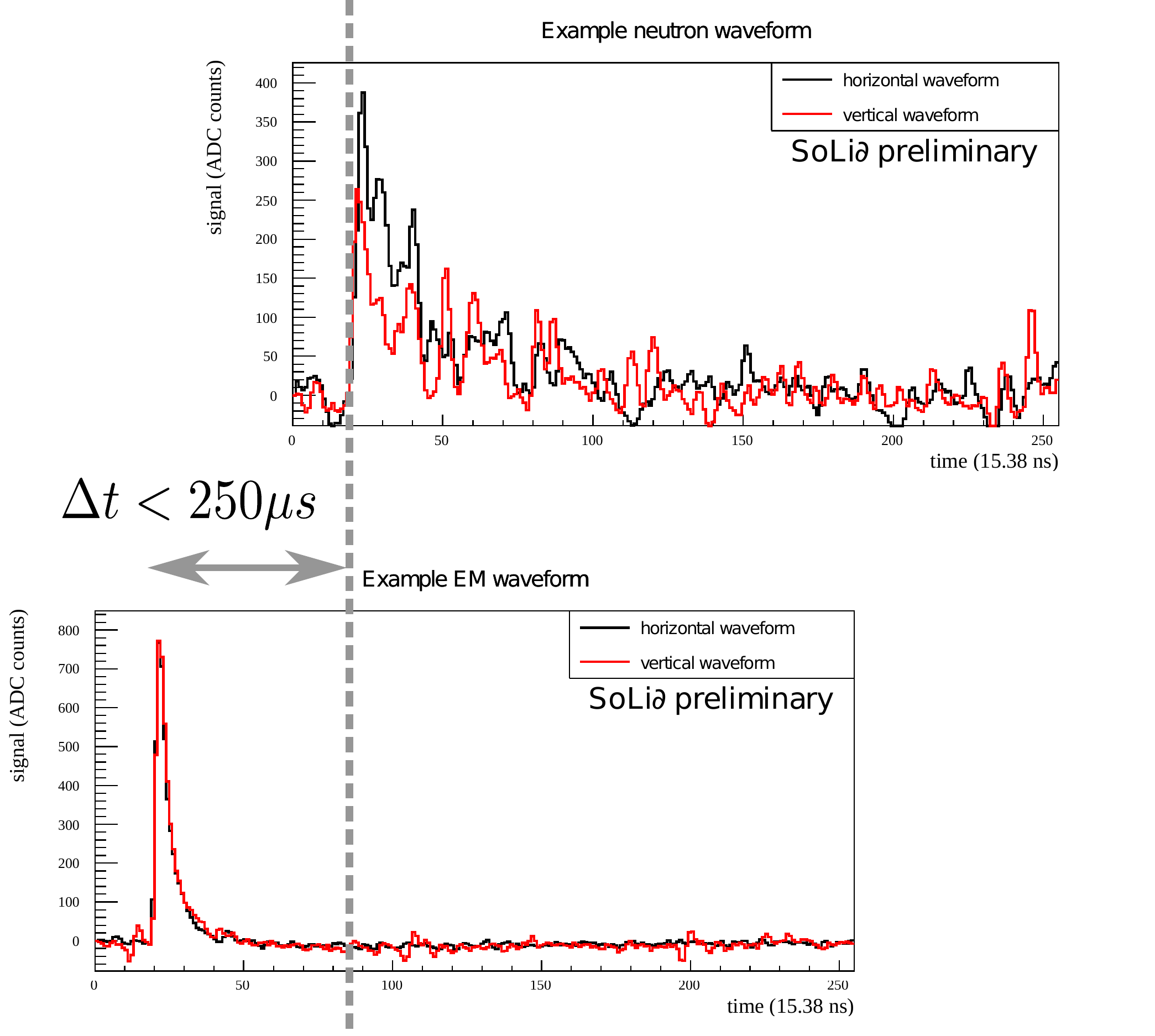}
        \caption{Example SiPM waveforms resulting from a positron (bottom) and neutron (top). Light is carried to the SiPMs by horizontal and vertical wavelength shifting fibres. The signals from the SiPM coupled to the horizonal(vertical) fibres are shown in black(red).}
        \label{fig:neutronEMwaveforms} 
    \end{subfigure}

    \caption{Illustration of inverse beta decay (IBD) in a SoLid scintillating cube and the signals resulting from positron and neutron. }
\end{figure}

The detector is segmented into optically isolated cubes of PVT scintillator.
Each cube is in contact with a thin sheet of $^6\mbox{LiF}$ mixed with ZnS(Ag) scintillator.
The positron from IBD is usually contained in a single cube.
The neutron thermalizes and can be absorbed by the $^6$Li which results in an $\alpha$ particle and a $^3$H nucleus with a combined energy of 4.8 MeV exciting scintillation in the ZnS(Ag), as illustrated in figure \ref{fig:compositescintillator}.
The light from the PVT and ZnS(Ag) scintillators have different decay times, allowing the positron and neutron signals to be separated. Example positron and neutron waveforms are shown in figure \ref{fig:neutronEMwaveforms}.
The $\gamma$ rays produced when the positron annihilates usually escape the cube where they are produced without depositing  a significant energy.
Hence the positron kinetic energy can be measured separately from the annihilation $\gamma$ ray energy.

\begin{figure}[htbp]
	\centering 
	\includegraphics[width=0.7\textwidth]{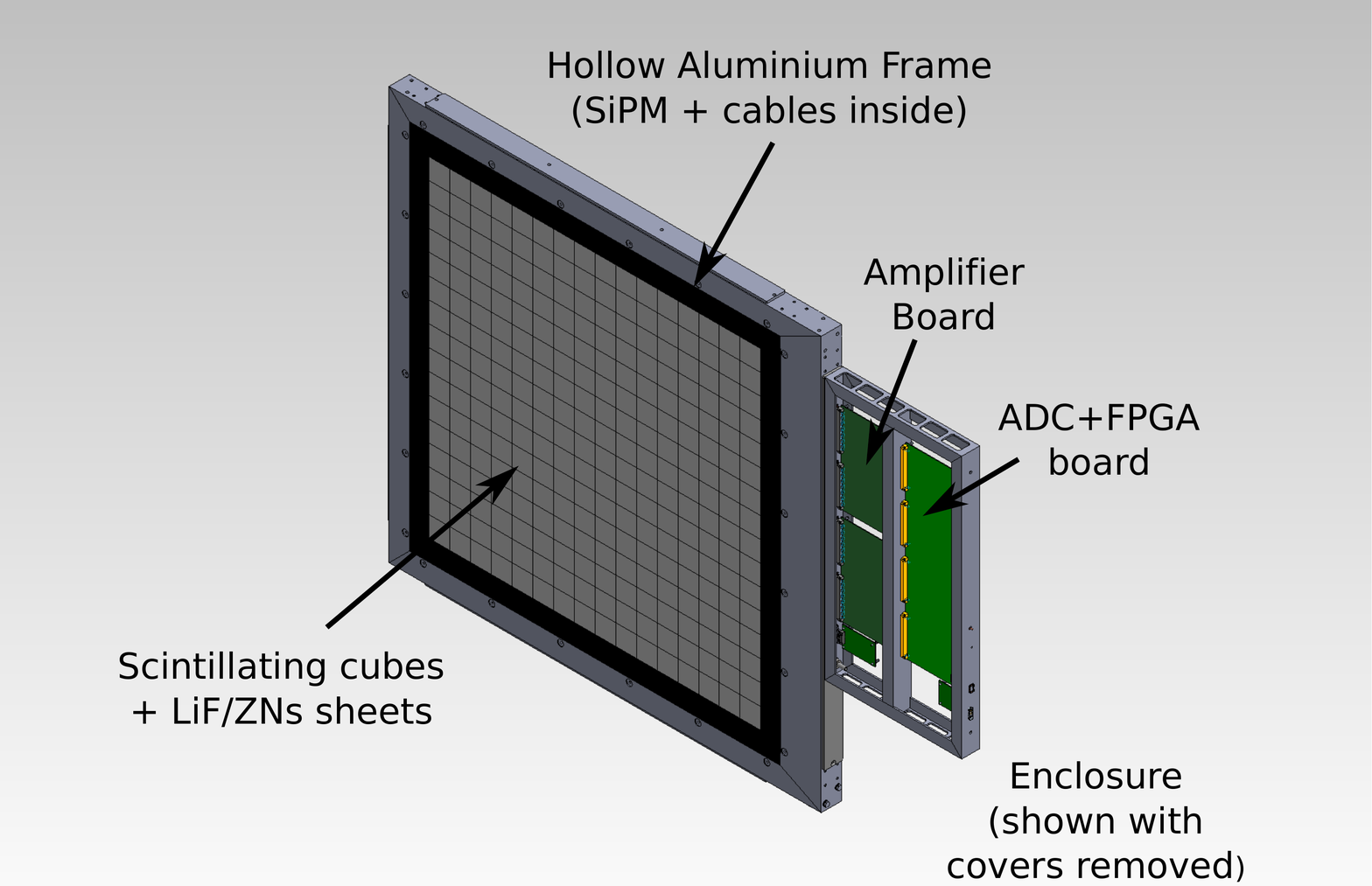}
	\caption{\label{fig:singleFrame} A 3D rendering of a single detector plane with readout electronics attached.}
\end{figure}

The cubes are arranged in planes containing 16 rows of 16 columns.
The light from each row and column of cubes is wavelength shifted and transported to SiPM photo-detectors by $3~\times~3$~mm$^2$ wavelength shifting fibres. There are two horizontal(vertical) fibres running through each row(column) of cubes. Hence the light from each cube is coupled to a total of four SiPM.
There are approximately 10 pixel avalanches (P.A.) in each SiPM per MeV of energy deposited in a cube, i.e. the total signal is $\approx$ 40 P.A./MeV \cite{ref:lalLightYield}.
An enclosure holding amplifier boards and a 64-channel ADC+FPGA board is attached to each detector plane.
Figure \ref{fig:singleFrame} is a 3D CAD rendering of a single frame with readout electronics attached.
The construction of the detector is described in more detail elsewhere \cite{ref:solidConstruction}. 


\begin{figure}[htbp]
	\centering 
	\includegraphics[width=0.7\textwidth]{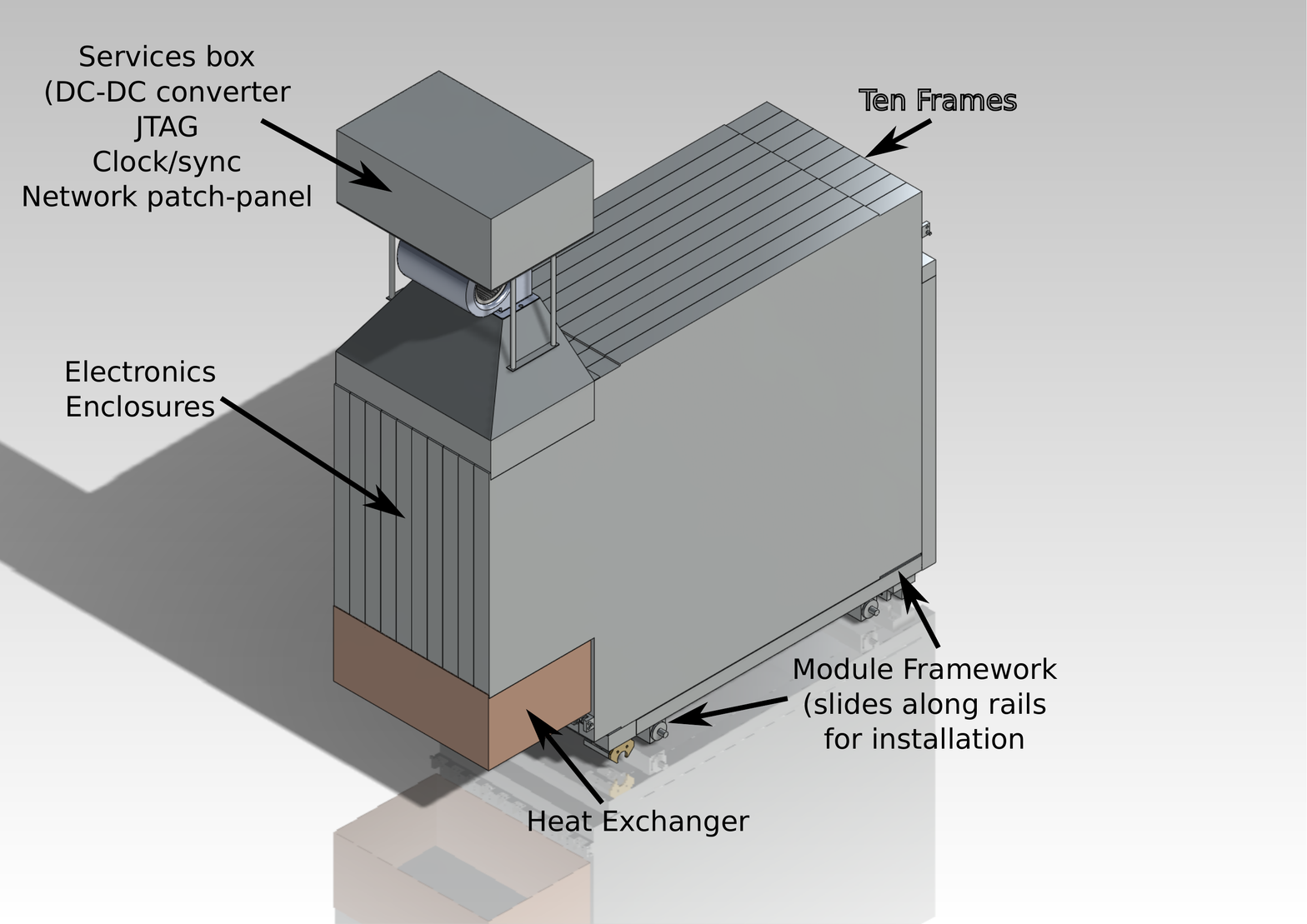}
	\caption{\label{fig:detectorModule} A 3D rendering of a ten plane detector module.}
\end{figure}

Ten detector planes, together with their readout electronics, form one detector module.
Detector modules are mechanically independent from each other and have separate power supply, clock and control distribution, cooling air blower and heat-exchanger.
Figure \ref{fig:detectorModule} shows a single module with the mechanical support, ten frames, blower, heat exchanger and services box for the module.
The services box contains converters, clock and synchronization distribution board, network patch panel and JTAG programming system.
Each module can be operated separately for commissioning, generating its own timing and control signals locally in the services box.
Modules are mounted on rails in a temperature stabilized container.

\FloatBarrier
\section{Readout hardware}
\label{sec:hardware}

\begin{figure}
	\centering
	\begin{minipage}{.6\textwidth}
		\centering
		\includegraphics[width=0.9\linewidth]{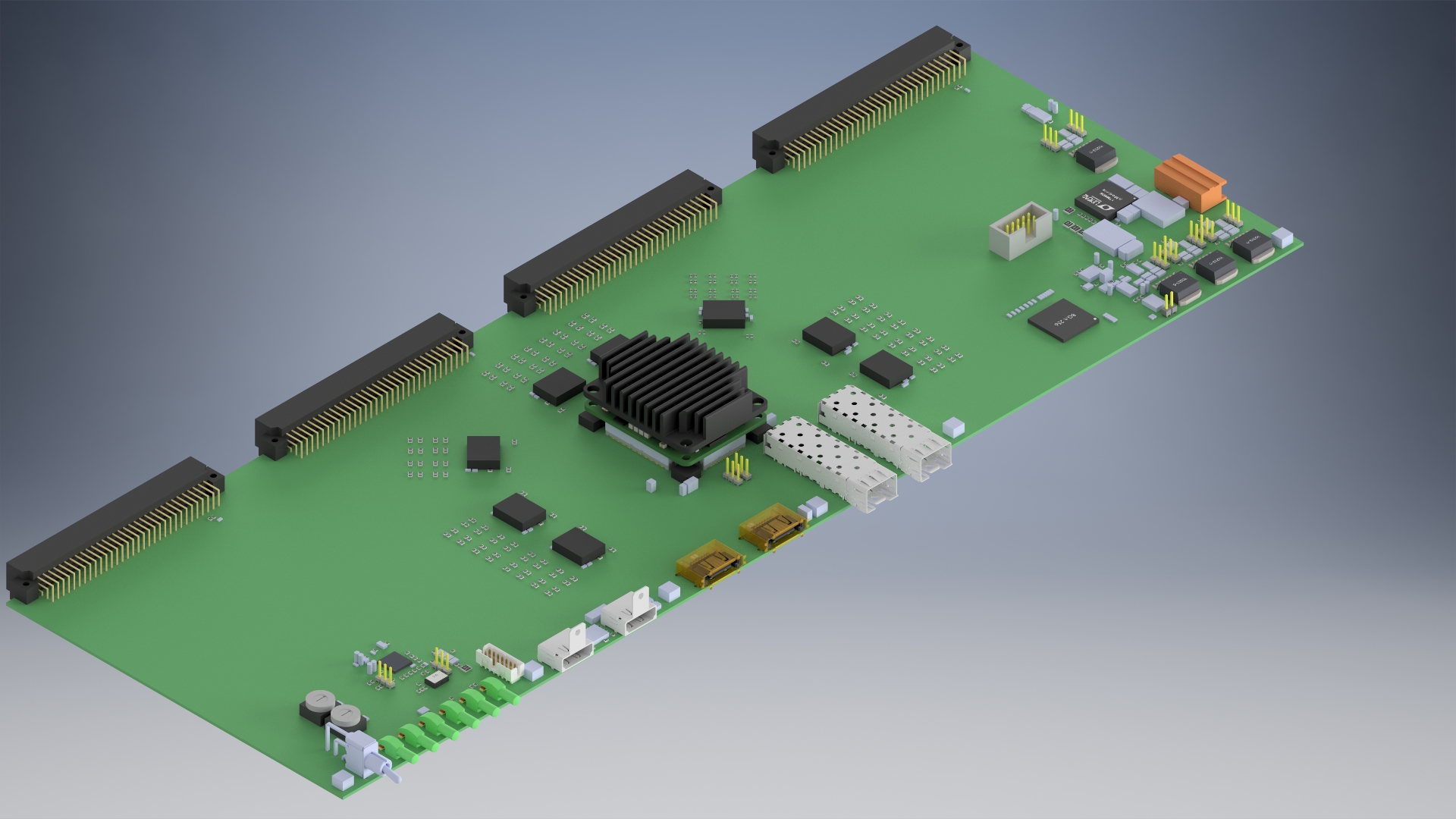}
		\captionof{figure}{3D CAD drawing of 64 channel ADC+FPGA board}
		\label{fig:64chanBoard}
	\end{minipage}%
	\qquad
	\begin{minipage}{.33\textwidth}
		\centering
		\includegraphics[width=.8\linewidth]{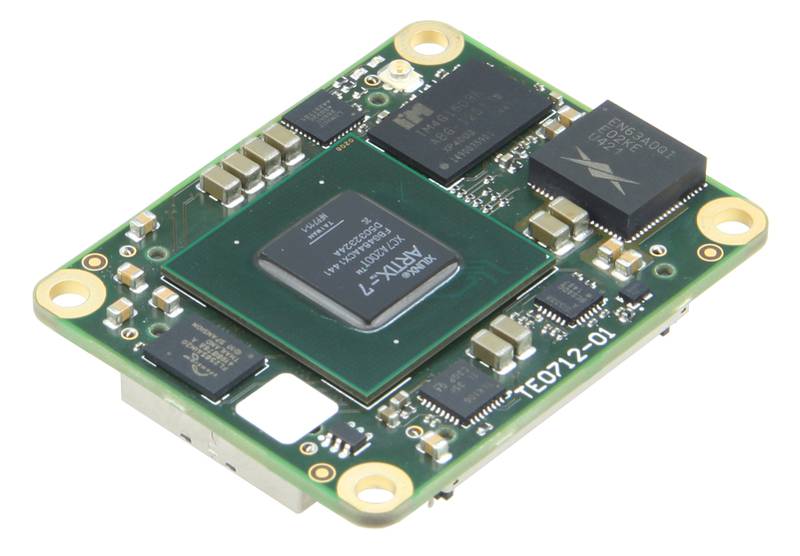}
		\captionof{figure}{Xilinx Artix-7 based FPGA board (Trenz TE0712)}
		\label{fig:te0712}
	\end{minipage}
\end{figure}

The SoLid experiment is aiming to rapidly deploy a novel detector with limited resources.
Given the significant number of channels that need to be instrumented and modest funding available it was not possible to use commercial off-the-shelf (COTS) readout modules.
Custom 64-channel readout boards have been designed using multiple 8-channel 14-bit 40MSample/s ADCs.
Readout over 1Gbit/s optical Ethernet is controlled by a COTS FPGA module.
Figure \ref{fig:te0712} shows the Trenz Xilinx Artix-7 based FPGA board used. 

\begin{figure}[htbp]
	\centering 
	\includegraphics[width=0.5\textwidth]{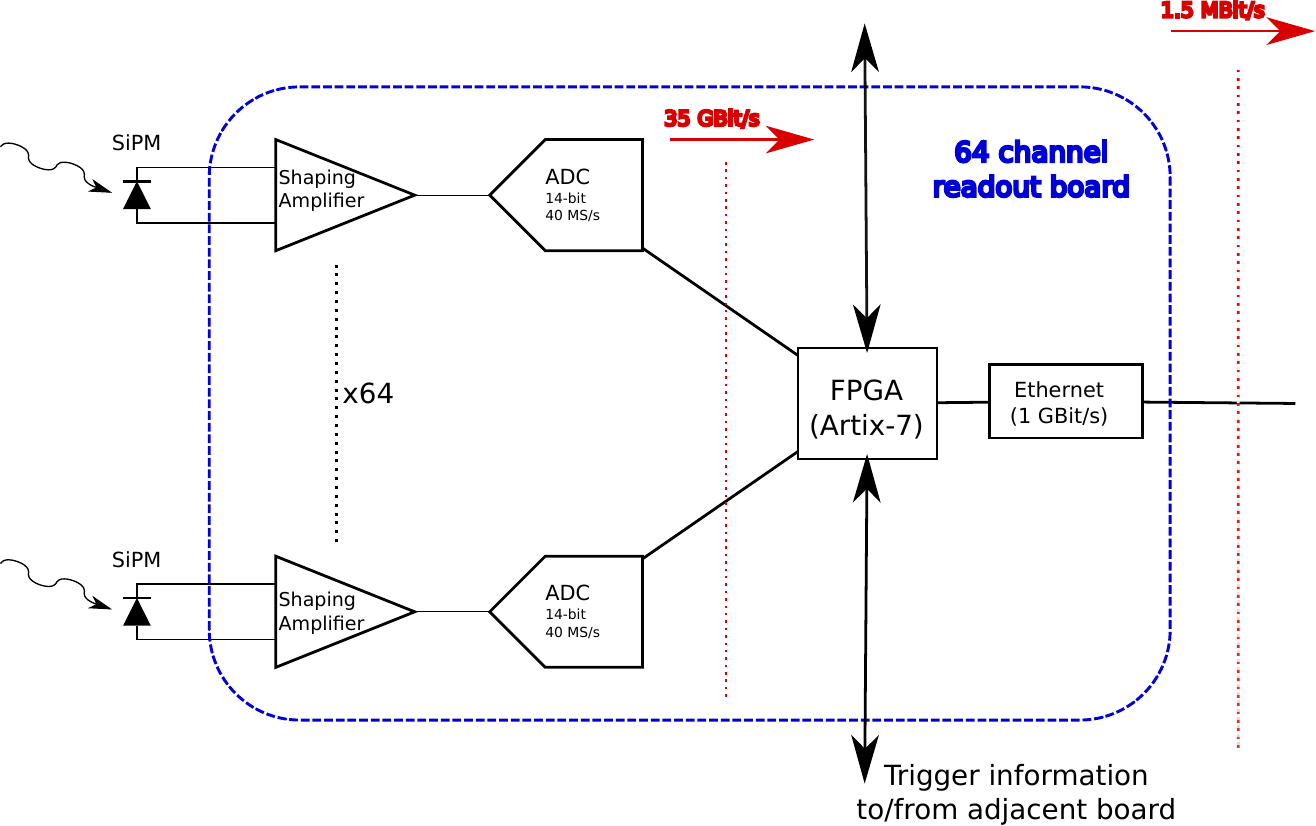}
	\caption{\label{fig:64ChanBlockDiagram} Block diagram of 64 channel readout hardware.}
\end{figure}

Each 64-channel ADC+FPGA readout board is attached to two 32-channel amplifier boards that apply a per-sensor programmable bias to the SiPM photo-detectors and amplify the signals from them.
The SiPMs are connected to the amplifier boards by twisted pair cables terminated into insulation displacement connectors.
The use of this twist and flat cable together with IDC connectors is cost effective and allows rapid connection to the photo-detectors.
Figure \ref{fig:64ChanBlockDiagram} is a block diagram of the 64-channel readout hardware associated with each detector plane.
Figure \ref{fig:64chanBoard} is a 3D rendering of a 64-channel readout board with ADC and FPGA board.
Figure \ref{fig:te0712} is a photograph of the Xilinx Artix-7 based COTS FPGA module.

The read out system is designed to have independent powering for each ten plane module.
Power is provided to the amplifier and readout boards at low voltage ( +5~V, -3.5~V and +5~V respectively).
Power converters in the services box generate these voltages from 48~V input.
There is additional common mode and differential power filtering at the entrance of each readout enclosure. 

\begin{figure}[htbp]
	\centering 
	\includegraphics[width=0.8\textwidth]{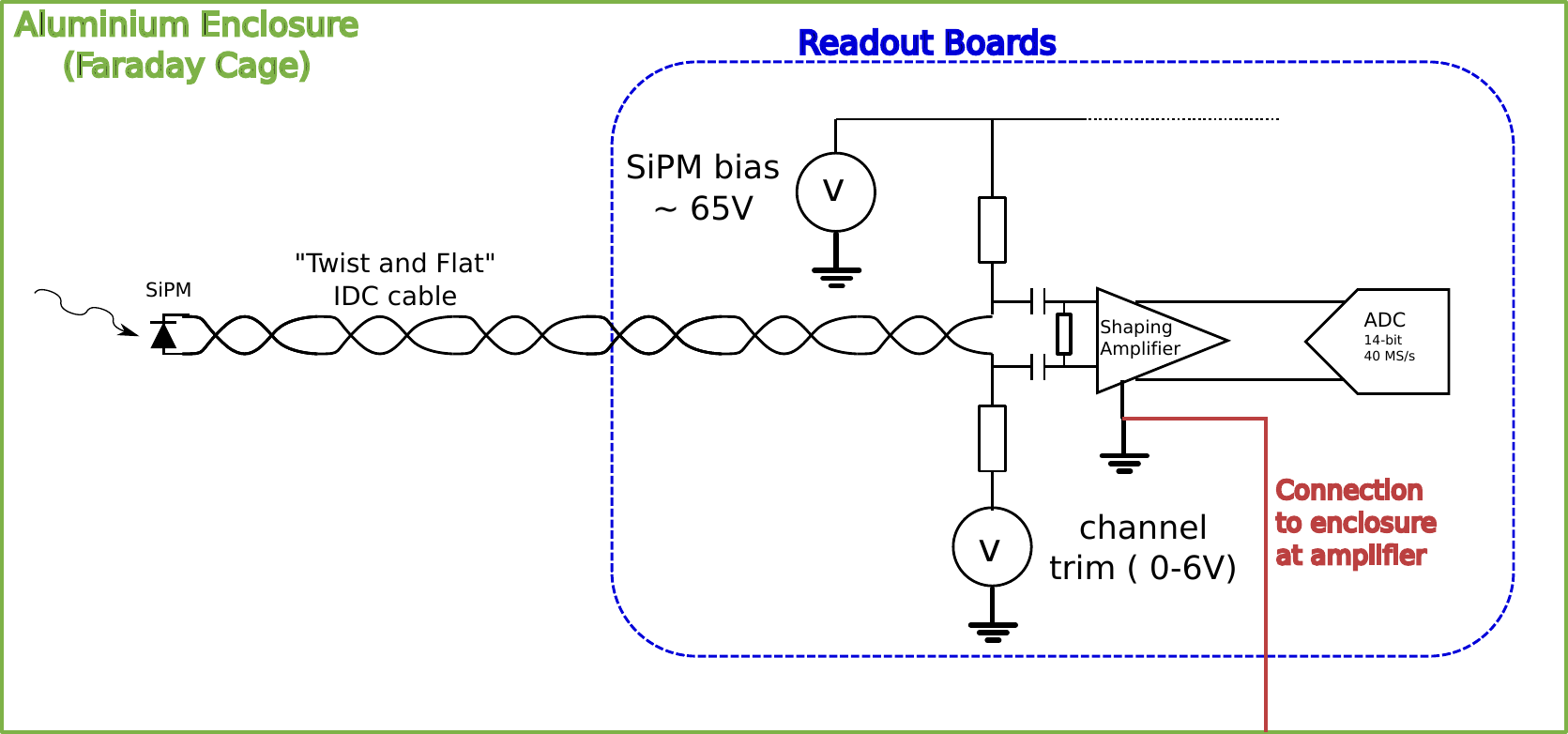}
	\caption{\label{fig:SolidGrounding} Sketch of grounding and shielding scheme for SoLid}
\end{figure}

The mechanical structure supporting the scintillating cubes is made of hollow extruded aluminium sections.
The sections are electrically connected together (the surfaces of the aluminium have a chromate conversion coating which makes the surface electrically conducting).
The enclosure for the readout electronics is electrically connected to the frame supporting the cubes using electrically conductive gaskets.
Hence each individual frame is a Faraday cage.
The signal reference of the amplifiers is connected at a single place to the readout enclosure.
Figure \ref{fig:SolidGrounding} shows a sketch of the grounding and shielding scheme.

\section{8-Channel Prototype}
\label{sec:8chan}

\begin{figure}[htbp]
	\centering 
	\includegraphics[width=0.8\textwidth]{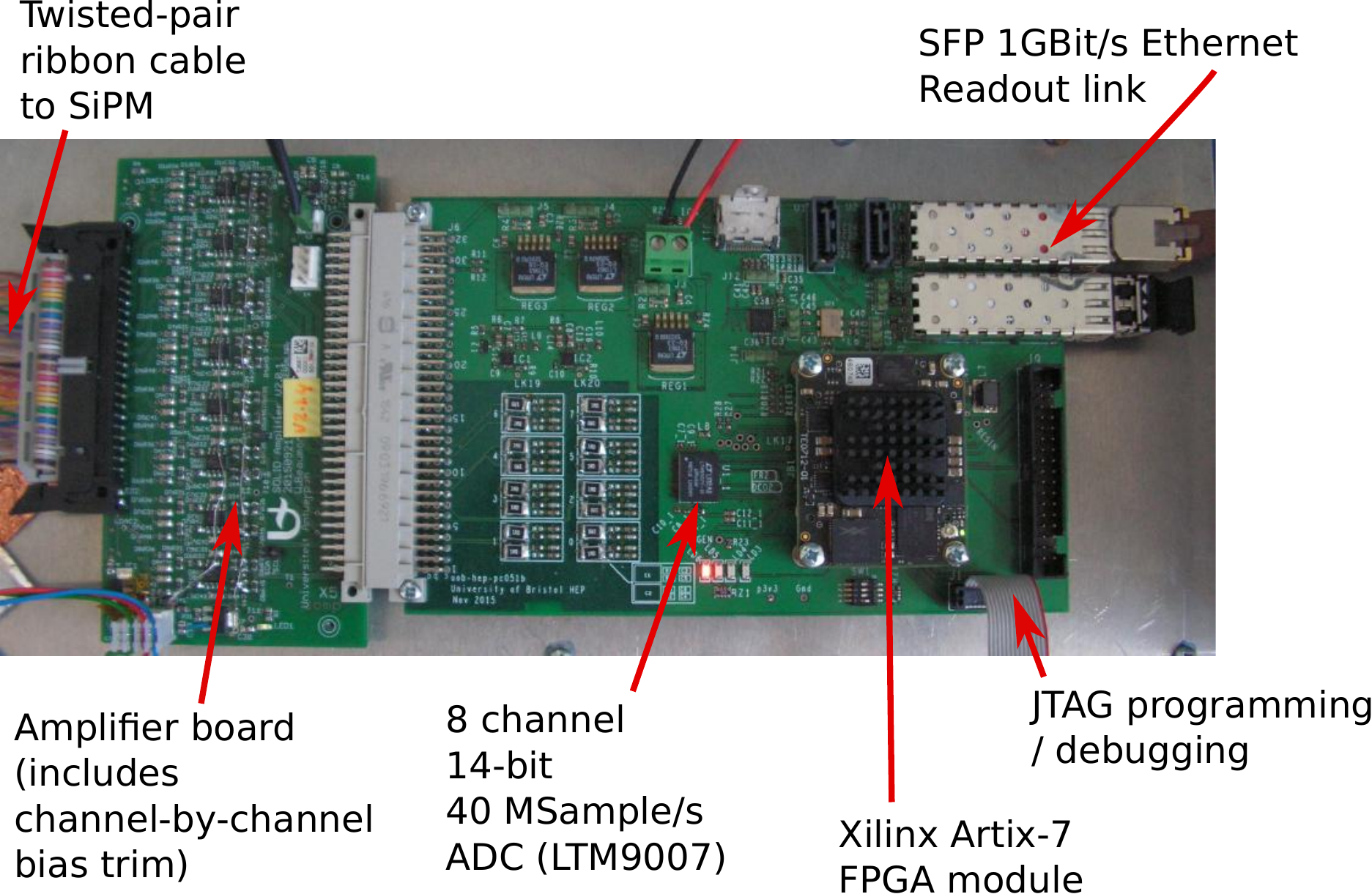}
	\caption{\label{fig:8ChanPhoto} The 8-channel prototype boards feature a single ADC. Key components are annotated.}
\end{figure}

\FloatBarrier
In order to test the readout concept an eight channel prototype using a single ADC was designed.
The 8-channel boards include all the major sub-systems present in the 64-channel boards, including clocking and synchronization circuitry and board to board Gbit/s links used to pass trigger information between boards.
These boards are also used for testing performance of detector prototypes.
Figure \ref{fig:8ChanPhoto} is an annotated photograph of an eight channel amplifier and ADC+FPGA readout board.

\begin{figure}[htbp]
    \centering
    \includegraphics[width=0.9\linewidth]{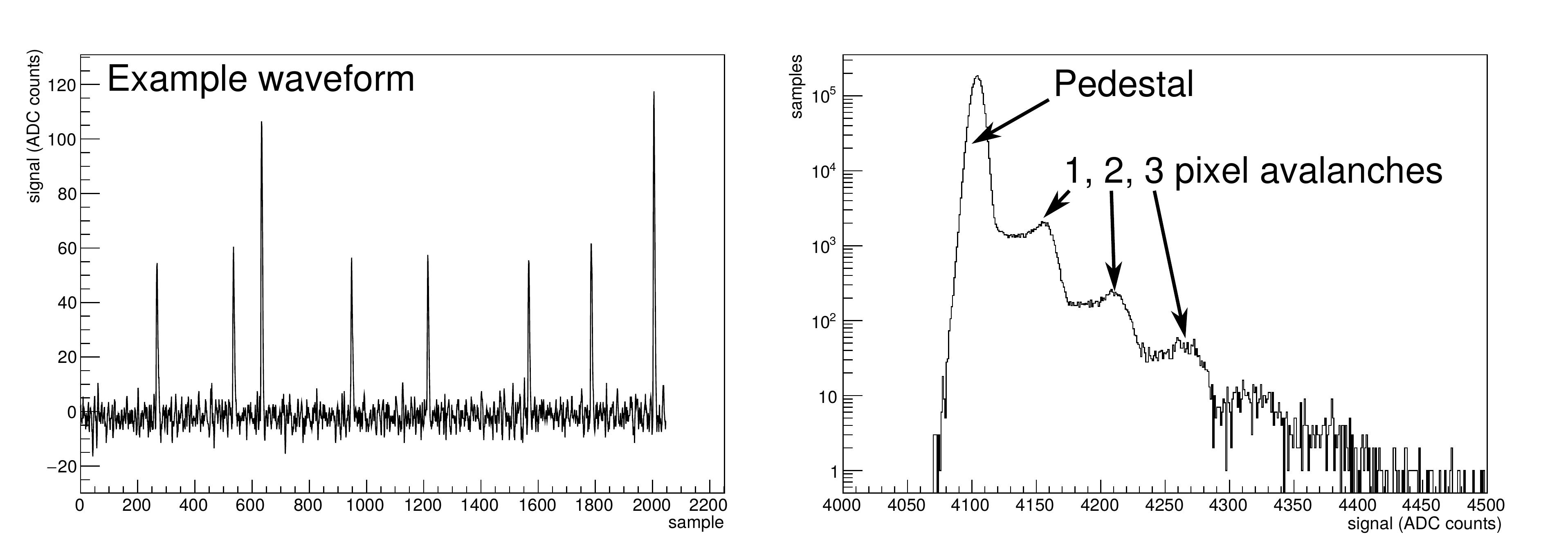}
    \captionof{figure}{An example waveform and the amplitude distribution of all samples from many such waveforms. Due to the low noise level peaks for 1, 2, 3 pixel avalanches are visible without any pulse selection.}
    \label{fig:8ChanPerformance}
\end{figure}

\begin{figure}[htbp]
    \centering
    \includegraphics[width=0.7\linewidth]{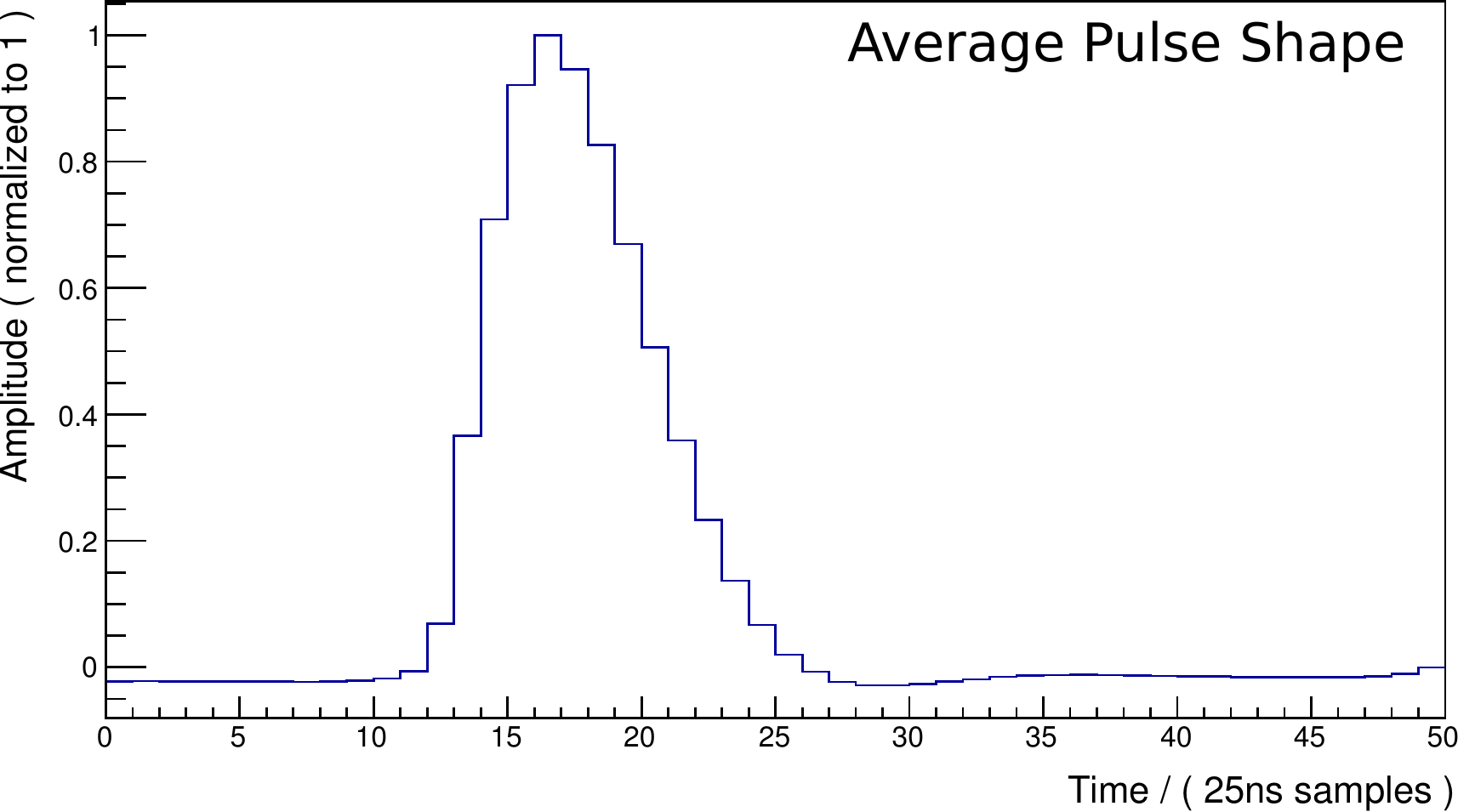}
    \captionof{figure}{Average pulse shape (normalized to an amplitude of one).}
    \label{fig:8ChanPulseShape}
\end{figure}

Initial tests show good performance.
Figure \ref{fig:8ChanPerformance} shows a typical waveform recording dark noise from a SiPM. 
It also shows the distribution of ADC output counts for all samples in a collection of such waveforms.
As can be seen, there is a noise level of a few ADC output counts, which is significantly less than the amplitude of a single pixel avalanche.
Figure \ref{fig:8ChanPulseShape} shows an averaged pulse shape for a SiPM signal.
With at least three samples on the rising edge and more on the falling edge it will be possible to perform a fit on the pulse shape to accurately determine the number of pixel avalanches in a pulse and its arrival time.


\FloatBarrier
\section{Firmware}
\label{sec:firmware}

\begin{wraptable}{r}{0.4\textwidth}
    \vspace{-6mm}
    \begin{tabular}{cc}
        \hline
        \hline
        Signal & hit rate \\
        \hline
        Dark count (SiPM) & GHz \\
        EM ($\gamma/\mu$) & 100 kHz \\
        neutron & Hz \\
        IBD & 0.05 Hz \\
        \hline
        \hline
    \end{tabular}
    \caption{Order of magnitude rates of different processes in SoLid.}
    \label{tab:eventRates} 
    \vspace{-5mm}
\end{wraptable}

\begin{figure}
	\centering
	\begin{minipage}{.35\textwidth}
		\centering
	\end{minipage}%
	\qquad
	\begin{minipage}{.55\textwidth}
		\centering

	\includegraphics[width=\textwidth]{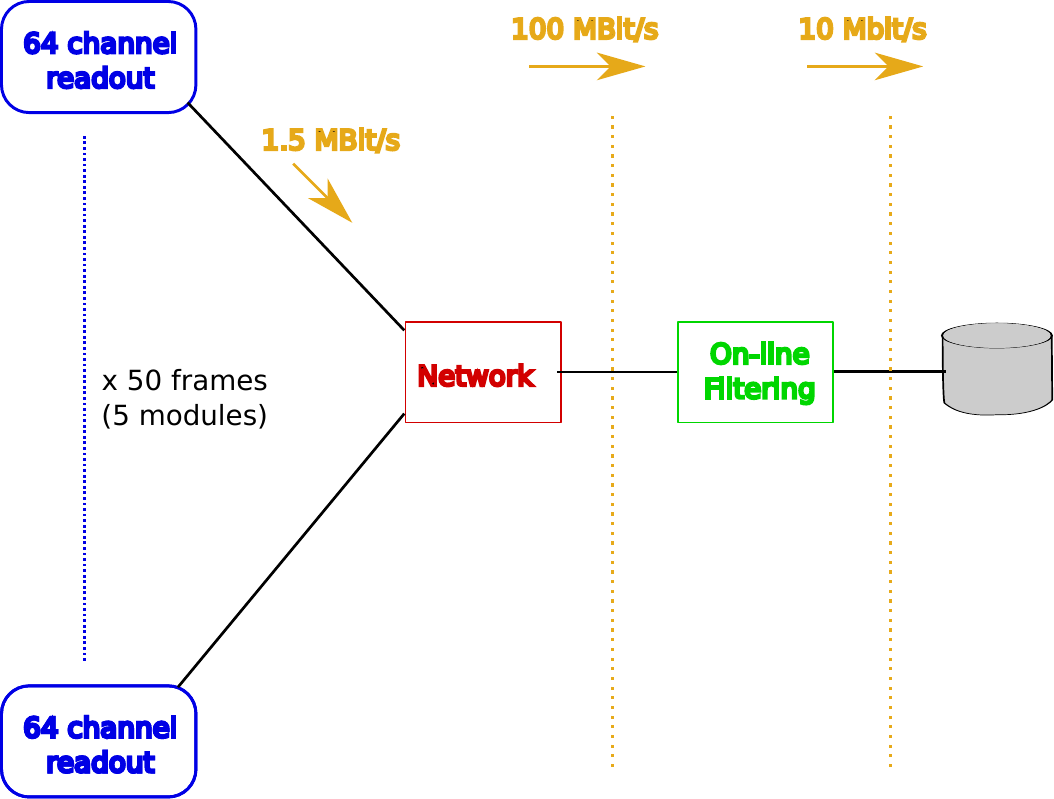}
	\caption{\label{fig:solidDataFlow} Block diagram data-flow through SoLid readout system.}
	\end{minipage}
\end{figure}

The 1600 kg detector to be deployed in 2017 has 3200 SiPM channels.
The ADCs digitise each channel with 14-bit resolution and a sampling rate of 40 MSample/s.
The rate of IBD neutrino interactions is $\approx 0.05 \mbox{Hz}$.
A high degree of on-line filtering is performed to reduce the data written to disk to O(10 Mbit/s).
Table \ref{tab:eventRates} lists the rates of the different processes in SoLid and figure \ref{fig:solidDataFlow} shows the data flow through the detector readout onto disk.

In order to cope with the degree of data reduction needed sophisticated firmware, employing both triggering and zero suppression, is required.
The trigger is based on detection of the neutron signal, which is independent of the neutrino energy, in order to avoid trigger bias distorting the measured neutrino energy spectrum.
When a neutron is detected approximately 1~ms of data centred on the trigger, which should contain the positron signal in an IBD event, is captured for readout. Events are read out over 1GBit/s Ethernet links using the IPBus protocol \cite{ref:IPBus} layered on top of UDP/IP.

\begin{figure}[htbp]
	\centering 
	\includegraphics[width=0.8\textwidth]{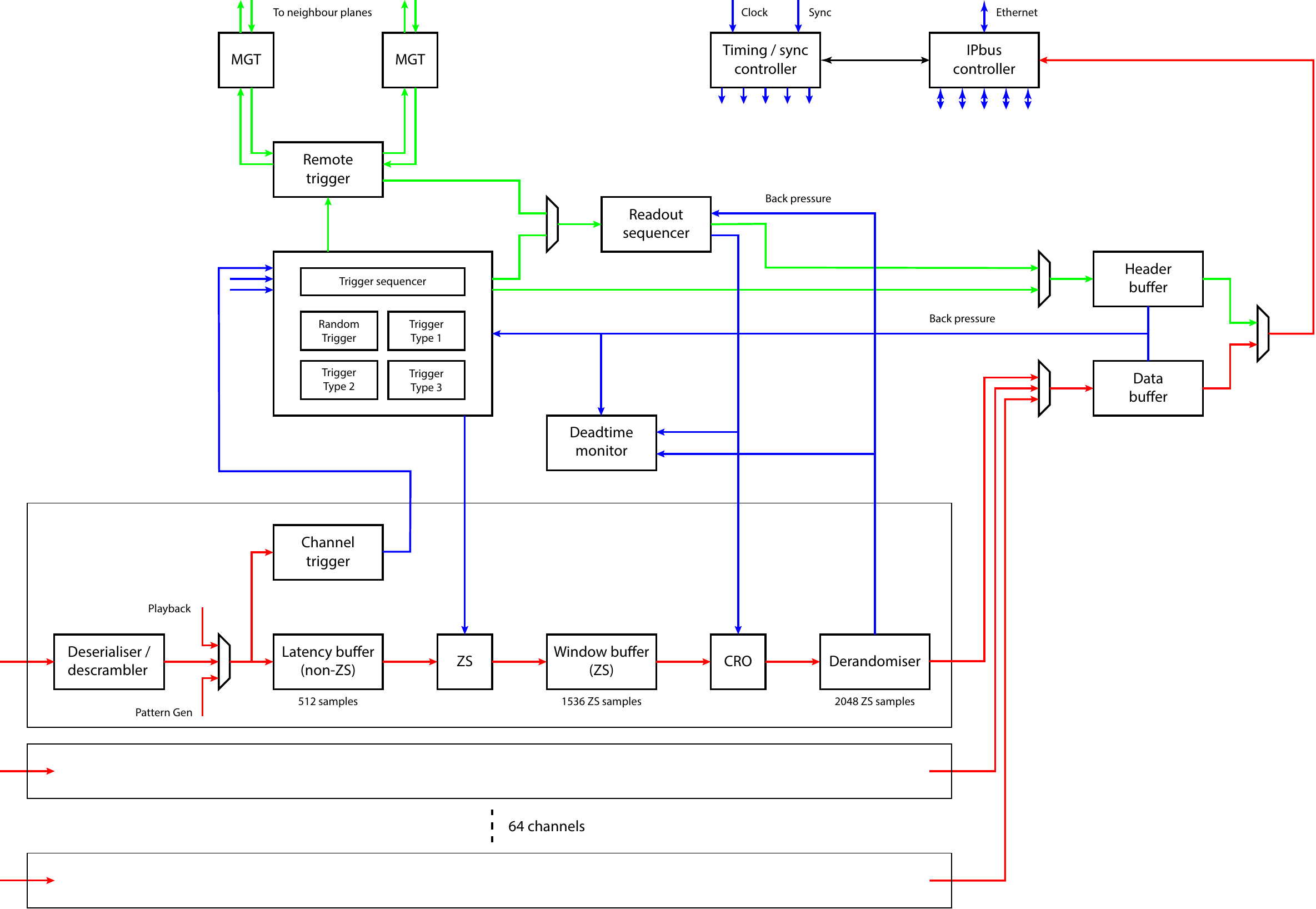}
	\caption{\label{fig:firmwareBlockDiagram} Block diagram of firmware}
\end{figure}

Figure \ref{fig:firmwareBlockDiagram} is a block diagram of the firmware.
Each ADC channel produces a 560 Mbit/s serial data stream that is decoded and put into a 512 sample ($12.8\mu\mbox{s}$) buffer.
At the end of the first latency buffer there is a zero suppression circuit.
Data is divided into blocks and if one or more samples is above threshold the block of data and its time of arrival are stored in a second zero suppressed buffer.
At the anticipated rate of dark noise the zero suppressed buffer will store approximately 1~ms of data.

In parallel with the first, non zero suppressed, latency buffer the neutron trigger logic operates on the data.
If the neutron trigger fires both the non zero suppressed data (which should contain the neutron signal) and the zero suppressed data ( which may contain data from the positron in an IBD event ) are stored for readout.

\begin{figure}[htbp]
	\centering
	\begin{minipage}{.35\textwidth}
		\centering
		\includegraphics[width=\linewidth]{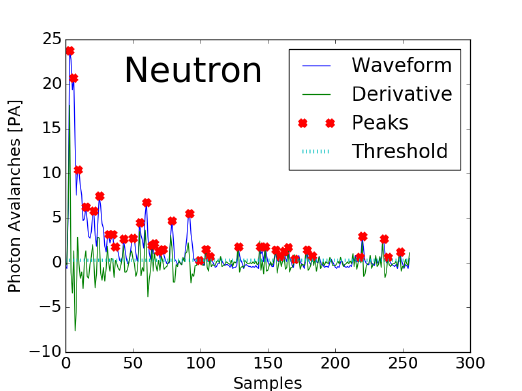}
		
		\includegraphics[width=\linewidth]{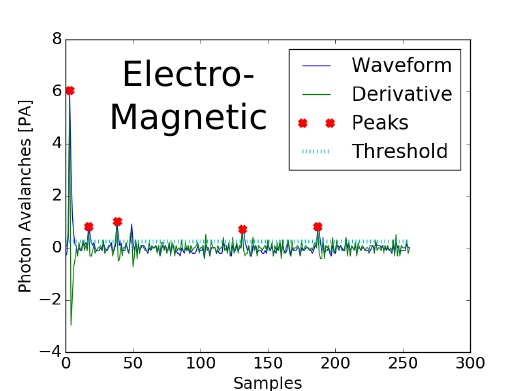}
		
		\caption{\label{fig:em_neutron_peaks} Typical SiPM output from ZnS (neutron) and PVT (EM) signals. Peaks from trigger firmware superimposed.}
		
	\end{minipage}%
	\quad
	\begin{minipage}{.6\textwidth}
		\centering
		\includegraphics[width=\linewidth]{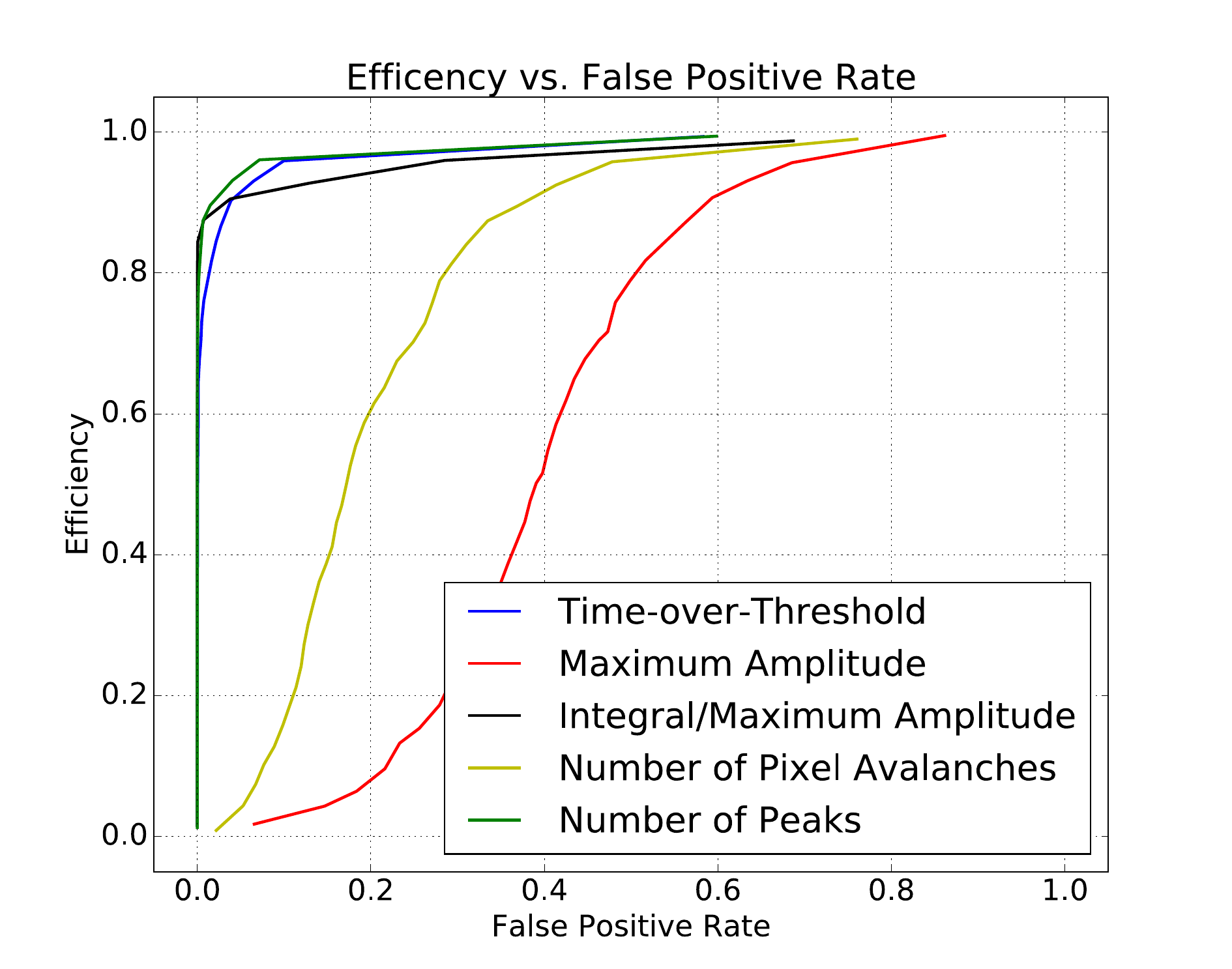}
        \captionof{figure}{Receiver Operator Curves for different neutron detection algorithms. The Time-over-Threshold and Number of Peaks algorithms considerably out-perform using a simple threshold trigger with offline neutron identification.}
		\label{fig:neutronTriggerROC}
	\end{minipage}
\end{figure}

The neutron trigger counts the number of peaks in a rolling time window.
As can be seen from figure \ref{fig:em_neutron_peaks}, the light from the ZnS(Ag) scintillator (which responds to neutrons) tends to produce many more peaks than the light from the PVT scintillator (which responds to EM induced energy deposits).
Counting peaks results in firmware that consumes relatively few FPGA resources and meets the desired level of trigger purity and efficiency, shown in figure \ref{fig:neutronTriggerROC}.
Trigger information is passed between detector frames and when a neutron trigger is generated data from the frames around the triggered frame are also read out.

\FloatBarrier
\section{Conclusion}


The SoLid collaboration have developed a highly modular readout system with a trigger scheme designed specifically for collecting data from candidate inverse beta decay events.
An FPGA-level algorithm for identifying neutron capture signals will be used to trigger data read out.
A millisecond scale FPGA-level data buffer is used to allow a time window read out of data from the region of interest around the neutron that is large enough to also contain positron signals from potential inverse beta decay events.
Prototype 8-channel versions of the full readout electronic boards have been produced and perform well.
The first test batch of the full scale electronics is currently in production, with a 3200 channel deployment at the BR2 reactor planned for the first half of 2017.

\acknowledgments
This work was supported by the following funding agencies: FWO-Vlaanderen and the Vlaamse Herculesstichting (Belgium); ANR, CEA and CNRS/IN2P3 (France); STFC (United Kingdom), DOE and NSF (United States). The research leading to these results has received additional funding from the European Research Council under the European Union's Horizon 2020 Programme (H2020-CoG) / ERC Grant Agreement n. 682474. We thank also our colleagues, the administrative and technical staffs of the SCK$\bullet$CEN for their invaluable support for this project. Individuals have received support from the FWO-Vlaanderen and the Belgian Federal Science Policy Office (BelSpo) under the IUAP network programme; The STFC Fellowship program; Merton College Oxford.


\end{document}